\documentstyle[11pt,newpasp,twoside,graphicx,psfig]{article}
\def\edcomment#1{\iffalse\marginpar{\raggedright\sl#1\/}\else\relax\fi}
\reversemarginpar

\begin{document}
\title{Young star clusters: Metallicity tracers in external galaxies}
\vspace{-0.4cm}
\author{Peter Anders, Uta Fritze - v. Alvensleben}
\affil{Universit\"ats-Sternwarte G\"ottingen, Geismarlandstrasse 11, 37083 G\"ottingen, Germany}
\author{Richard de Grijs}
\affil{Department of Physics \& Astronomy, The University of Sheffield, Hicks Building, Hounsfield Road, Sheffield S3 7RH, UK}

\begin{abstract}
Star cluster formation is a major mode of star formation in the extreme conditions of interacting galaxies and violent starbursts. These newly-formed clusters are built from recycled gas, pre-enriched to various levels within the interacting galaxies. Hence, star clusters of different ages represent a fossil record of the chemical enrichment history of their host galaxy, as well as of the host galaxy's violent star formation history.
We present a new set of evolutionary synthesis models of our {\sc galev} code, specifically developed to include the gaseous emission of presently forming star clusters, and a new tool to analyse multi-color observations with our models. First results for newly-born clusters in the dwarf starburst galaxy NGC 1569 are presented.
\end{abstract}

\vspace{-0.4cm}
\section{Models \& tests}
We use our evolutionary synthesis code {\sc galev} to study the basic physical parameters of star clusters in interacting galaxies and violent starbursts. Star clusters can easily be approximated as simple stellar populations (SSPs), since all stars have the same age, metallicity and extinction. However, more complicated star formation / metallicity evolution histories can be studied by superimposing appropriate SSPs. The main ingredients of our code are: stellar isochrones (ensemble of stars with different masses at a given age) from the Padova group for metallicities in the range of -1.7 $\le$ [Fe/H] $\le$ +0.4, a stellar initial mass function (usually assumed to be Salpeter-like), the library of stellar spectra by Lejeune et al. (1997,1998), and gaseous emission (both lines and continuum). From the integrated spectra we derive magnitudes in a large number of filter systems. The gaseous line and continuum emission contributes significantly to the integrated light of stellar populations younger than $3 \times 10^7$ yr both in terms of absolute magnitudes and derived colors (Anders \& Fritze - v. Alvensleben 2003). The updated models are available from \begin{center} {\bf http://www.uni-sw.gwdg.de/$\sim$galev/panders/} \end{center}

\begin{figure}
\begin{center}
\includegraphics[angle=-90,width=0.75\linewidth]{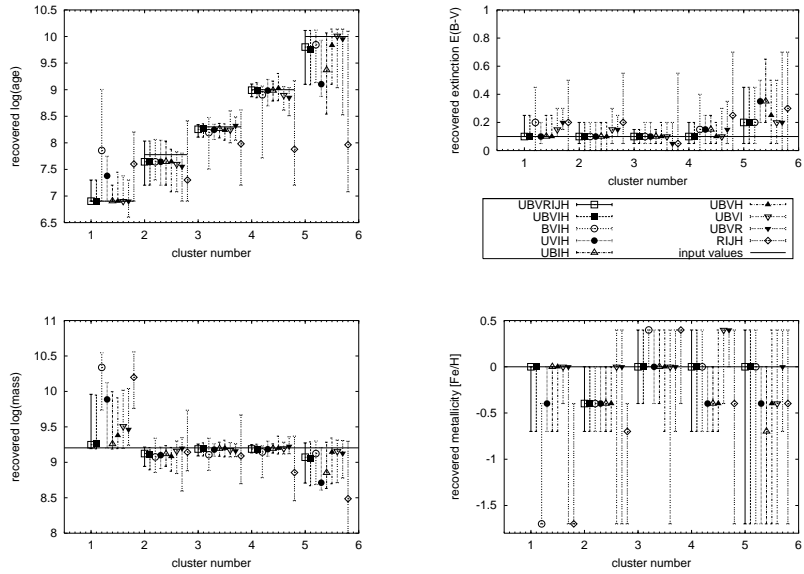}
\end{center}
\caption{Dispersion of recovered properties of artificial clusters, assuming availability of {\sl UBVRIJH} and passband combinations with 4 out of the available 7 passbands, as indicated in the legend. Cluster parameters are: solar metallicity, E(B-V) = 0.1 mag, assumed ``observational'' error = 0.1 mag, ages = 8, 60, 200 Myr, 1, 10 Gyr.}
\end{figure}

\begin{figure}
\begin{center}
\includegraphics[angle=-90,width=0.75\linewidth]{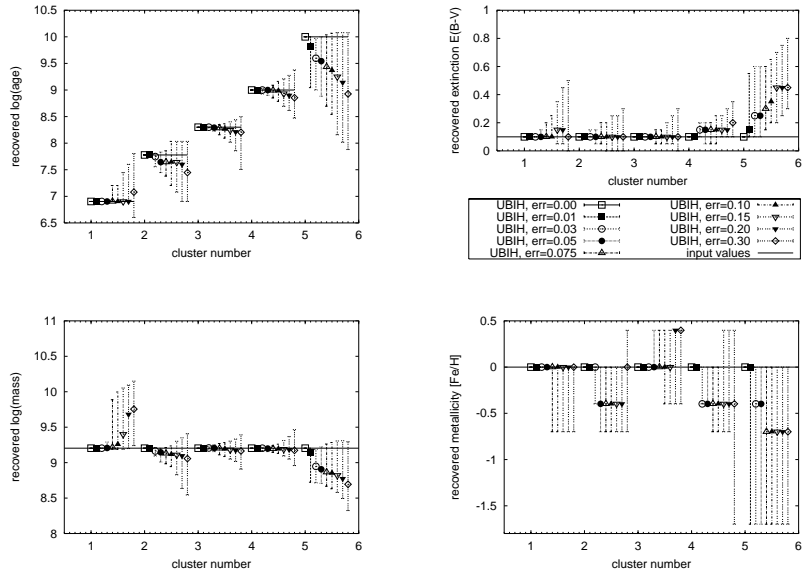}
\end{center}
\caption{Dispersion of recovered properties of artificial clusters, using the best 4-passband-combination {\sl UBIH}. Cluster parameters are: solar metallicity, E(B-V) = 0.1 mag, ages = 8, 60, 200 Myr, 1, 10 Gyr. The assumed ``observational'' errors are indicated in the legend.}
\end{figure}

We developed a tool to compare our evolutionary synthesis models with observed cluster SEDs to determine the basic cluster parameters age, metallicity, internal extinction and mass independently. Using artificial clusters with various input parameters (with cluster SEDs taken directly from the evolutionary synthesis models) we systematically studied the impact of the choice of passbands, of finite observational magnitude uncertainties, and {\sl a priori} assumptions. Two examples of the artificial cluster tests are shown in Fig. 1 and 2. Additional tests were performed using broad-band observations of star clusters in NGC 3310 (de Grijs et al. 2003a,b), confirming the results from the artificial cluster tests. Due to the young age of this cluster system these additional tests are restricted to ages younger than approx. 200 Myr. From these tests we conclude that:
\begin{enumerate}
\item At least 4 passbands are necessary to determine the 3 free parameters age, metallicity and extinction, and the mass by scaling the SED, independently.
\item The most important passbands are the {\sl U} and {\sl B} bands; for systems older than roughly 1 Gyr the {\sl V} band is equally important.
\item NIR bands significantly improve the results by constraining the metallicity efficiently.
\item A wavelength coverage as long as possible is desirable. Best is UV through to NIR, thus tracing the pronounced kink/hook in the SEDs around the {\sl B} band.
\item Large observational errors and/or wrong {\sl a priori} assumptions may lead to completely wrong results.
\end{enumerate}

\vspace{-0.5cm}
\section{The case of NGC 1569}
As a first application the dwarf starburst galaxy NGC 1569 was chosen. Our sample enlarges the number of star clusters studied in this galaxy by a factor of 3. Our results for the derived physical parameters are in agreement with previous results of the starburst history in NGC 1569 in general, and of the two prominent ``super star clusters'' in particular, regarding age, mass and metallicity. In addition, we find a surprising change in the cluster mass function with age: The clusters formed during the onset of the burst (approx. 25 Myr ago) seem to exhibit an excess of massive clusters as compared to clusters formed more recently. Using various statistical methods we show the robustness of this result (Anders et al. 2003b).

\vspace{-0.5cm}
\section{References}
Anders, P., Fritze - v. Alvensleben, U. 2003, \aap, 401, 1063\\
Anders, P., Bissantz, N., Fritze - v. Alvensleben, U., de Grijs, R. 2003a,\\ \mnras, {\sl submitted}\\
Anders, P., de Grijs, R., Fritze - v. Alvensleben, U., Bissantz, N. 2003b,\\ \mnras, {\sl submitted}\\
de Grijs, R., Fritze - v. Alvensleben, U., Anders, P., Gallagher, J. S., Bastian, N., Taylor, V. A., Windhorst, R. A. 2003a, \mnras, 342, 259\\
de Grijs, R., Anders, P., Bastian, N., Lynds, R., Lamers, H.J.G.L.M., O'Neil, Jr., E.J. 2003b, \mnras, 343, 1285\\
Lejeune, T., Cuisinier, F., Buser, R. 1997, \aaps, 125, 229\\
Lejeune, T., Cuisinier, F., Buser, R. 1998, \aaps, 130, 65
\end{document}